\documentclass[epsfig,aps,prl,twocolumn]{revtex4}
\usepackage{graphicx}
\usepackage{amsmath}
\usepackage{revsymb}
\usepackage{natbib}

\def\r{\rangle}

\begin{document}
\bibliographystyle{apsrev}

\title{Multi-Photon Entanglement Concentration and Quantum Cryptography }

\author{Gabriel A. Durkin$^1$, Christoph Simon$^1$, and Dik Bouwmeester$^{1,2}$}

\affiliation{$^1$ Centre for Quantum Computation, University of Oxford, Oxford OX1 3PU, United Kingdom \\
$^2$ Department of Physics, Center for Spintronics and Quantum Computation, University of California, Santa
Barbara, CA 93106, USA}
\date{\today}

\begin{abstract} Multi-photon states from parametric down-conversion can be entangled both
in polarization and photon number. Maximal high-dimensional entanglement can be concentrated post-selectively from
these states via photon counting. This makes them natural candidates for quantum key distribution, where the
presence of more than one photon per detection interval has up to now been considered as undesirable. We propose a
simple multi-photon cryptography protocol for the case of low losses.
\end{abstract}

\maketitle

Parametric down-conversion is a convenient way of creating entangled states of light \cite{bouwmeester}. So far it
has been explored in two separate regimes. Experiments on the few-photon level have often relied on polarization
entanglement \cite{kwiatPDC}, while experiments with macroscopic amounts of light have observed two-mode squeezing,
that is entanglement in photon number \cite{kimble}. It is however possible to build sources that combine both
kinds of entanglement. The basic principle has recently been demonstrated \cite{lamas}. Such a source can be seen
as a pair of phase-coherent two-mode squeezers. We will show that from this point of view photon counting can be
used as a post-selective realization of entanglement concentration for continuous variable states. Maximal
high-dimensional entanglement can be extracted in this way from the multi-photon states.

It is natural to consider the application of this entanglement for quantum key distribution. For the original
quantum cryptography protocols \cite{bb84,ekert} the presence of more than one photon in a single pulse or
detection interval is a problem for security. Therefore implementations of key distribution \cite{entpairs} are
usually restricted to weak transmission signals, with a low probability of containing even a single photon,
limiting the achievable bit rate per pulse. The pulse rate itself is mainly limited by the dead time of the photon
detectors. Here we take a more positive approach to multi-photon states in cryptography. We ask whether they can be
used to improve the capacity of the secure channel. We propose a simple protocol which leads to a significant increase in
bit rates for the case of low losses.

We will first describe our proposed post-selective realization of entanglement concentration for
continuous-variable states. Entanglement concentration is a procedure that allows two parties Alice and Bob to
extract maximal entanglement from non-maximally entangled pure states using only local operations and classical
communication \cite{bennett}. Consider the (un-normalized) two-mode squeezed state
\begin{equation}
|\psi_1\r=\sum_{l=0}^{\infty} \lambda^l |l \rangle_{a_h} |l
\rangle_{b_v} \label{non-max}
\end{equation}
where $\lambda$ is usually referred to as the squeezing parameter. For later convenience, we have assumed that the
photons in the spatial mode $a$ (going to Alice) are horizontally and those in mode $b$ (going to Bob) are
vertically polarized. This state represents photon-number entanglement between modes $a_h$ and $b_v$, that is, a
quantum superposition of different states for which the number of photons in mode $a_h$ and $b_v$ are the same. The
state is however not maximally entangled since $\lambda$ is always smaller than unity and therefore the individual
terms in the superposition have different weights.

Based on ref. \cite{duan} we describe a way to concentrate photon-number entanglement. Suppose that in addition to
(\ref{non-max}) Alice and Bob are also given the state $|\psi_2\r=\sum_{m=0}^{\infty} (-\lambda)^m |m
\rangle_{a_v} |m\rangle_{b_h}$, which differs from (\ref{non-max}) by the sign of the squeezing parameter and by
the polarization of the photons in modes $a$ and $b$. The total state is then given by:
\begin{equation}
|\Psi\rangle=|\psi_1\r|\psi_2\r=\sum_{l=0}^{\infty}\sum_{m=0}^{\infty}\lambda^{l+m} (-1)^m
|l\rangle_{a_h}|m\rangle_{a_v}|m\rangle_{b_h}|l\rangle_{b_v} \,.
\label{phase}
\end{equation}
Defining $n=l+m$, rearranging terms and using the short-hand
notation $|u,v;w,x\rangle$ for $|u\rangle_{a_{h}}|v\rangle_{a_{v}}|w\rangle_{b_{h}}|x\rangle_{b_{v}}$ yields
\begin{equation}
|\Psi\rangle =
\sum_{n=0}^{\infty}\lambda^n\left(\sum_{m=0}^{n}(-1)^m
|(n-m),m;m,(n-m)\rangle\right),
\label{entangle}
\end{equation}
where we have collected the terms with the same number of photons $n$ received by Alice and Bob.

Entanglement concentration could now be achieved by performing a projection measurement onto a specific photon
number. For a given $n$, this results in a superposition state of $n+1$ equally weighted terms. It is evident from
$(\ref{entangle})$ that each term satisfies $(N_v-N_h)_a=(N_h-N_v)_b$. From our subsequent discussion it will
become clear that these perfect correlations in photon number difference exist not only in the h/v polarization basis, but in {\it any}
basis. This is a consequence of the maximal entanglement in (\ref{entangle}), i.e. of the fixed phase relations
between the different $|(n-m),m;m,(n-m)\rangle$ terms.

At first glance, the above scheme seems to require a quantum non-demolition (QND) measurement of the photon number
on each side in order to project onto a fixed value of $n=(N_v+N_h)_a=(N_v+N_h)_b$ without losing the possibility
of measuring $(N_v-N_h)_a$ and $(N_h-N_v)_b$ afterwards. Ways of realizing such a QND measurement were discussed in
\cite{duan}, but it is very difficult to implement. On the other hand, {\it destructive} photon counting is
feasible. It is therefore important to realize that for many applications it is not strictly necessary to perform
the $n$ projection before the $(N_v-N_h)_a$ and $(N_h-N_v)_b$ measurements. They can be performed simultaneously by
simply measuring $(N_v)_a$, $(N_h)_a$, $(N_h)_b$ and $(N_h)_b$ independently. The basis of polarization analysis
can be varied, permitting the observation of perfect correlations in more
than one basis. This approach is similar to the post-selection
strategy that enabled the demonstration of quantum teleportation \cite{postselection} and related single-photon
experiments.

Clearly, one should be careful in referring to a post-selection method as a concentration scheme since no
concentrated output state is obtained. However, for the purpose of quantum cryptography the post-selection method
will suffice, since it allows to establish perfect correlations between Alice's and Bob's measurement results.

In quantum key distribution, to prevent eavesdropping it is essential that perfect correlations are obtained in at
least two {\it complementary} bases. If there were perfect correlations only in one basis, the eavesdropper could
make her measurements in this basis, and the process would not be secure. We now show that, due to our specific
choice of relative phases, the state (\ref{entangle}) is symmetric under a joint rotation of polarisation bases
through equal angles in modes `a' and `b'. Therefore, the state exhibits the same photon-number difference
correlations in, for example, the linear polarization basis rotated by $45^{\circ}$. We also show how such a
symmetric state (\ref{entangle}) can be generated in a natural way using type-II parametric down-conversion.

Parametric down-conversion is a process where a photon from a pump light source can be split into two photons of
lower frequency within a non-linear optical crystal. One can experimentally achieve conditions where a good
approximation for the relevant interaction Hamiltonian is
\begin{eqnarray}
\hat{H}=\kappa (\hat{a}^{\dagger
}_{h}\hat{b}^{\dagger }_{v}-\hat{a}^{\dagger }_{v}\hat{b}^{\dagger }_{h})  + h.c.,
\label{Hamiltonian}
\end{eqnarray}
where the complex number $\kappa$ is the product of the amplitude of the pump beam and the relevant non-linear
coefficient of the crystal. This is the familiar Hamiltonian for the creation of polarization entangled photon
pairs \cite{kwiatPDC}, which has been the basis for many experiments in quantum information. Using the normal
ordering theorem of \cite{normalorder} one can show that this Hamiltonian leads to the production of entangled
photon states of the following form:
\begin{eqnarray}
|\psi\rangle &=& \exp (-i \hat{H} t /\hbar)|0\rangle \nonumber\\ &=& \frac{1}{\cosh^{2}(\tau)}\sum
_{n=0}^{\infty}\sqrt{n+1}
\;\tanh^{n}(\tau)\;|\psi^{n}_{-}\rangle, \label{pdcstate}
\end{eqnarray}
where $\tau=\frac{\kappa t}{\hbar }$ is the effective interaction time and
\begin{eqnarray}
|\psi^{n}_{-}\rangle &=& \frac{1}{\sqrt{n+1}}
\frac{1}{n!}(\hat{a}^{\dagger }_{h}\hat{b}^{\dagger
}_{v}-\hat{a}^{\dagger }_{v}\hat{b}^{\dagger }_{h})^n |0\rangle \nonumber\\
&=&
\frac{1}{\sqrt{n+1}}\sum_{m=0}^{n}(-1)^{m}|(n\!\!-\!\!m),m;\;\!m,(n\!\!-\!\!m)\rangle
\,.
\end{eqnarray}
The total state (\ref{pdcstate}) has exactly the form of state (\ref{entangle}). The terms $|\psi^{n}_{-}\rangle$,
which correspond to $n$ photons on each side, are maximally entangled states shared between Alice and Bob in a
Hilbert space of $(n+1)\times(n+1)$ dimensions. Similar states were studied in the context of Bell's inequalities
in \cite{drummond}. They are all invariant under joint identical polarization transformations by Alice and Bob,
since they are created by the application to the vacuum of powers of the operator $(\hat{a}^{\dagger
}_{h}\hat{b}^{\dagger }_{v}-\hat{a}^{\dagger }_{v}\hat{b}^{\dagger }_{h})$, whose form is conserved under such
transformations. These properties make them generalized singlet states, which motivates our notation $|\psi_-^n\r$.
Whenever Alice has $(n-m)$ photons polarized along a certain direction and $m$ photons polarized along the
orthogonal one, Bob has $m$ and $(n-m)$ photons of the respective polarizations. When employed for quantum key
distribution, every pair of values $(m,n-m)$ constitutes a letter in the cryptographical alphabet.

A simple key distribution protocol using the multi-photon states proceeds in the following way. From a common
source, entangled multi-photon pulses are sent to Alice and Bob via modes $a$ and $b$. Alice and Bob each
independently and randomly choose one of two complementary bases, h/v and h'/v', in which to perform their photon number measurements. Here, the primed basis is rotated by $45^{\circ}$ with respect to the unprimed basis. These measurements act as a post-selective multi-photon
entanglement concentration resulting in detected correlations associated with the states $|\psi_-^n\r$, where $n$
is the number of detected photons on each side. They communicate their basis choice via classical means and extract
the key from the photon number difference recorded in those cases where they had chosen the same basis. Finally,
Alice and Bob examine a randomly chosen part of the key for errors. In the ideal case any amount of errors
indicates the presence of an eavesdropper.

It is clear that in the absence of losses the achievable bit rate increases significantly with the number of
photons because the number of distinguishable measurement outcomes increases. There are $n+1$ different possible
measurement results for the state $|\psi_-^n\r$. For protocols based on the multi-photon states $|\psi^n_-\rangle$,
photon losses introduce errors because the state after losses no longer has the perfect correlations expected. We
will model photon losses by the action of beam-splitters introduced to each of the four modes
($a_{h},a_{v},b_{h},b_{v}$). The probabilities for the measurement of particular photon numbers in each mode can be calculated using a positive operator valued measure (POVM). The operator associated with a
measurement of `n' photons in mode $a_{h}$ (behind the beam-splitter) is
\begin{eqnarray}
\hat{\mathbf{P}}_{n}= \eta^n \sum_{m=0}^{\infty}\frac{(m+n)!}{m!n!}(1-\eta)^m |n+m\rangle_{a_{h}} \langle n+m|
\label{POVM}
\end{eqnarray}
Here, $\eta$ is the transmission coefficient of the beam-splitter, and corresponds to the overall
quantum efficiency of the system, including lossy lines and imperfect detectors. Each term in (\ref{POVM})
corresponds to a certain number of photons $m$ that were lost. We assume the same amount of loss in all four photon
modes. Probabilities of specific outcomes are calculated by taking the expectation value of the associated POVMs
with the down-conversion state Eq.(\ref{pdcstate}). Thus the probabilities are functions only of $\eta$ and $\tau$.

The information shared between Alice and Bob can be quantified by the
\emph{mutual information} \cite{mutualinf}:
\begin{eqnarray}
I_{AB}=\frac{\sum_{A,B}p(A_{i},B_{j})log_{2}p(A_{i},B_{j})}{\sum_{i}p(A_{i})log_{2}p(A_{i})\sum_{i}p(B_{i})log_{2}p(B_{i})},
\end{eqnarray}
which is a function of the joint probabilities for Alice's and Bob's measurement results, denoted by $A_i$ and
$B_j$ respectively. An outcome labelled $A_{i}$ corresponds to a particular pair of photon-numbers measurement made
on Alice's side; it will be of form $(n-k)$ photons in mode $a_h$ ($a_{h'}$), and $k$ photons in mode $a_v$ ($a_{v'}$), where the basis
of polarization analysis depends on her choice.

In quantum cryptography, Alice and Bob have to assume that all errors that seem to be due to losses could actually
be the consequence of eavesdropping, with the eavesdropper Eve simulating the effect of lossy lines. In such a
situation, Eve will have some knowledge about Alice's and Bob's results, quantified by the mutual informations
$I_{AE},I_{BE}$.  In the presence of an eavesdropper, the number of secure shared bits that Alice and Bob can distill by privacy amplification techniques \cite{privamp} is denoted the `secrecy capacity' $C_s$ , and is limited by the
inequality \cite{ekert2}:
\begin{equation}
C_s \geq I_{AB}-\mbox{min}(I_{AE},I_{BE}).
\label{cs}
\end{equation}
Determining the achievable secure bit rates in principle requires an analysis of all possible eavesdropping
strategies. This is a difficult task in the present situation since the system under consideration is very complex.
In this paper, as a first step, we consider a specific key distribution protocol where Alice and Bob make use of
the 4-photon detection results (each detects 2 photons) in addition to the 2-photon results (each detects 1
photon). We have compared this case to the standard protocol which exclusively uses the 2-photon results
\cite{entpairs}. As for the eavesdropping strategy, we suppose that Eve's technology is so powerful that she can
replace the lossy transmission lines, unknown to Alice and Bob, by ideal ones. Furthermore, we assume that Eve
controls the source. She is aware that Alice and Bob will monitor errors, and tailors her interference to reproduce
the error profiles expected. Indeed, there are two types of errors that Alice and Bob can check.

The first type is the occurrence of photon number detections different from one on each side or two on each side.
These results are produced under normal circumstances, despite not being used for key generation. Eve has no choice
but to replicate these signals, labelled below in Eq.(\ref{evesource}) as $\hat{\rho}_{\text{rest}}$.

The second type of error occurs when Alice and Bob both measure the same number of photons, but not the ideal
perfect correlations in polarization. Eve sends 2-photon and 4-photon signals with the expected overall
probabilities, $P_{1,1}(\eta,\tau),P_{2,2}(\eta,\tau)$, but does not always send the singlet states,
$|\psi_{1}^{-}\rangle, |\psi_{2}^{-}\rangle$, which give her no information, and give Alice and Bob perfect
correlations. Instead, a proportion of the time defined by $\gamma$, Eve sends a product state with the correct
correlations in one basis. She has no way of knowing the basis, $\oplus$ (h/v) or $\otimes$
(h'/v') in which the legitimate users will measure, and is forced to choose randomly. If she guesses
correctly, she has full knowledge of their results. However, when her basis choice differs from that of Alice and
Bob, she introduces correlation errors on their measurements. The percentage $\gamma$ is constrained to produce
exactly the frequency of natural errors expected on the 2- and 4-photon signals. Therefore it is also a function of
$\eta$ and $\tau$. The state produced by Eve's source is:
\begin{equation}
\hat{\rho}_{Eve}=\text{P}_{1,1}\hat{\rho}_{1,1}+\text{P}_{2,2}\hat{\rho}_{2,2}+(1-\text{P}_{1,1}-\text{P}_{2,2})\hat{\rho}_{\text{rest}} \label{evesource}
\end{equation}
where for instance:
\begin{widetext}
\begin{eqnarray}
\hat{\rho}_{1,1}=(1-\gamma)|\psi_{1}^{-}\rangle \langle \psi_{1}^{-}| \nonumber + \frac{\gamma}{4}
(|1,0;0,1\rangle_{\otimes}\langle 1,0;0,1 |+|0,1;1,0\rangle_{\otimes}\langle 0,1;1,0 |+   \nonumber
|1,0;0,1\rangle_{\oplus}\langle 1,0;0,1 |+|0,1;1,0\rangle_{\oplus}\langle 0,1;1,0 |) \\
\end{eqnarray} 
\end{widetext}
The subscripts $\oplus, \otimes$ label the two complementary polarisation bases in which each product
state is defined. The state $\hat{\rho}_{2,2}$ is defined analogously. From an explicit description of the full
state as given above one can directly calculate the joint probabilities for all possible measurement outcomes,
which determine each mutual information and thus the minimum secrecy capacity.
\begin{figure}[ht!]
\includegraphics[width=3.5in]{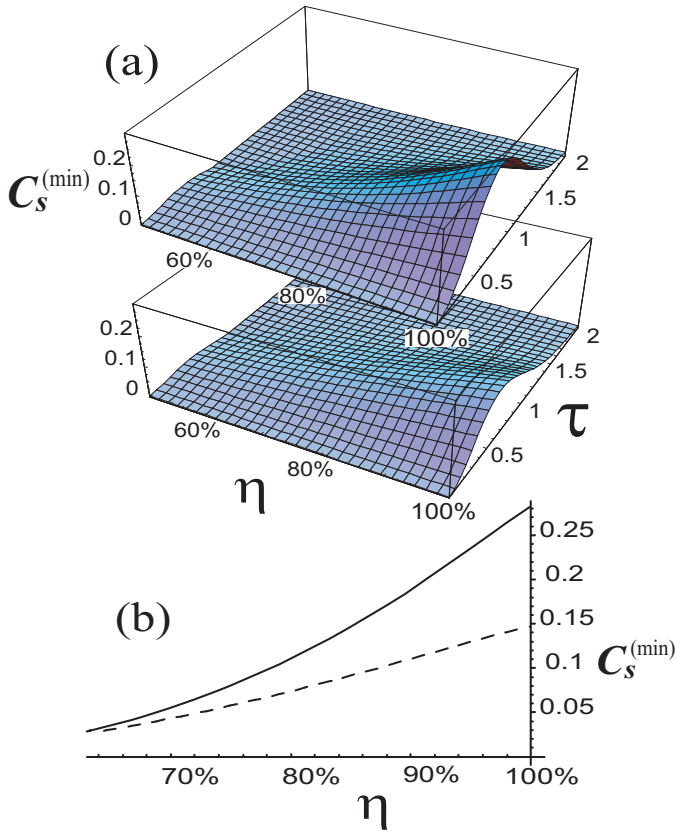}
\caption{\small{The minimum achievable secure bit rate $C_s^{\text{(min)}}$ Eq. (\ref{cs}) for the \emph{multi-photon}
protocol is depicted in (a) upper graph, in the context of the specific eavesdropping attack mentioned in the text.
To contrast, the equivalent measure for the standard protocol, using 2-photon results only, is shown in (a), lower
graph. Graphs are plotted in terms of the overall transmission $\eta$, and the effective interaction time of the
source $\tau$, cf. eq. (\ref{pdcstate}). One sees that using 4-photon detections in addition leads to a significant
increase in secure bit rates in the region of low losses. This is shown in more detail in (b) where we have plotted
$C_s^{\text{(min)}}$ for both protocols, at their optimal $\tau$ values; $\tau=0.78$ and $\tau=0.70$ for the
multiphoton and standard protocols respectively. $C_s^{\text{(min)}}$ decreases for higher $\tau$ values, as can be
seen clearly in (a), because the probabilities for 2-photon and 4-photon results are reduced as higher photon
numbers become more likely.}}
\label{results}
\end{figure}

The results are shown in figure \ref{results}. One sees that for a comparatively low level of losses the minimum
secrecy capacity is approximately doubled by using the 4-photon states in addition. This effect would be increased
substantially by including higher photon numbers.

It should be noted in this context that efficient multi-photon detectors \cite{kim} and optical fibres with very
low losses \cite{fibres} are both under development. Currently, losses and limited detection efficiencies are
serious practical restrictions. One can see from fig. \ref{results}(b) that for the present protocol the advantage
of using the higher photon number states disappears for overall losses that exceed 35 \%. However, there is
some indication that the multi-photon states may still be viable candidates for quantum key distribution for higher
losses. The entanglement in the states $|\psi_{n}^{-}\rangle$ is quite robust under photon loss. We will address
this topic in a future publication. The entanglement that remains after some loss could be purified and then used
for key distribution or other quantum communication tasks.

Natural applications for multi-photon entanglement include all-optical quantum error correction \cite{dikec} and
even all-optical quantum computation \cite{knill}. The use of down-conversion multi-photon states for these
purposes is a topic for future research.

We thank G. Giedke and L. Vaidman for stimulating discussions. This work was supported by the EPSRC GR/M88976 and
the European Union QuComm (IST-1999-10033) projects.

\end{document}